\documentclass[preprint]{imsart}

\RequirePackage{amsthm,amsmath,amsfonts,amssymb}
\RequirePackage[authoryear]{natbib}
\RequirePackage[colorlinks,citecolor=blue,urlcolor=blue]{hyperref}

\usepackage{enumitem,graphicx,color,verbatim,multirow,pgf,pgfarrows,pgfnodes}
\usepackage{tikz}
\usetikzlibrary{arrows.meta}
\usepackage[font=small,format=plain,labelfont=bf,up,textfont=normal,up,justification=justified,singlelinecheck=false]{caption}
\usepackage{array,longtable,tabu}
\newcolumntype{^}{>{\currentrowstyle}}

\usepackage{xcolor}
\usepackage{comment}

\theoremstyle{definition}
\newtheorem{theorem}{Theorem}

\newcommand{\Diag}{\operatorname{Diag}}

\begin{document}
\begin{frontmatter}

\title{Probabilistic HIV Recency Classification \,---\, A Logistic Regression without Labeled Individual Level Training Data}
\runtitle{Probabilistic HIV Recency Classification}

\begin{aug}
\author[A]{\fnms{Ben} \snm{Sheng}},
\author[A]{\fnms{Changcheng} \snm{Li}},
\author[A]{\fnms{Le} \snm{Bao}\ead[label=e1]{lebao@psu.edu}},
\author[A]{\fnms{Runze} \snm{Li}}
\address[A]{Department of Statistics, Penn State University, University Park, PA, USA\\
\printead{e1}}
\runauthor{Sheng et. al.}
\end{aug}

\begin{abstract}
Accurate HIV incidence estimation based on individual recent infection status (recent vs long-term infection) is important for monitoring the epidemic, targeting interventions to those at greatest risk of new infection, and evaluating existing programs of prevention and treatment. Starting from 2015, the Population-based HIV Impact Assessment (PHIA) individual-level surveys are implemented in the most-affected countries in sub-Saharan Africa. PHIA is a nationally-representative HIV-focused survey that combines household visits with key questions and cutting-edge technologies such as biomarker tests for HIV antibody and HIV viral load which offer the unique opportunity of distinguishing between recent infection and long-term infection, and providing relevant HIV information by age, gender, and location.
In this article, we propose a semi-supervised logistic regression model for estimating individual level HIV recency status. It incorporates information from multiple data sources \,---\, the PHIA survey where the true HIV recency status is unknown, and the cohort studies provided in the literature where the relationship between HIV recency status and the covariates are presented in the form of a contingency table. It also utilizes the national level HIV incidence estimates from the epidemiology model. Applying the proposed model to Malawi PHIA data, we demonstrate that our approach is more accurate for the individual level estimation and more appropriate for estimating HIV recency rates at aggregated levels than the current practice \,---\, the binary classification tree (BCT). 
\end{abstract}

\end{frontmatter}

\section{Introduction}
\label{sec:Intro}

HIV incidence is the rate of new infections in a susceptible population during some time period (e.g., 1 year), and specifies the current state of the epidemic.  It is perhaps the most important indicator for critical health policies regarding monitoring the epidemic, targeting interventions to those at greatest risk of new infection, and evaluating existing programs for prevention and treatment.  Therefore, estimating incidence, especially at small-area level, is crucial for an effective and efficient response to the HIV epidemic \citep{Hallett2011, WHO2020}.  Unfortunately, HIV incidence estimation is challenging because HIV infections last a lifetime but initially show few or nonspecific symptoms, so many newly infected individuals do not get tested and diagnosed until months or years after infection \citep{Jones2019, Giguere2021}. In HIV testing, the individual level HIV recency status (recent vs long-term infection) is also much harder to evaluate than the HIV infection status (positive vs negative).  A number of methodologies have been developed to make inference for HIV incidence at national or sub-national levels.  Population-level HIV incidence could be modeled indirectly from facility-level HIV surveillance data using the Spectrum-EPP epidemic models, but this requires some strong assumptions about unobserved parts of the epidemic \citep{Bao2012rtrend,Bao2010sti,Bao2012modelling,Brown2010,Brown2014,Hogan2010}. As another approach, HIV incidence could be measured more directly from incidence assays \citep{Welte2009,Kassanjee2012}. However, the required sample size for accurate estimates would be impractically large since recent HIV infections are relatively rare in the susceptible population \citep{Bao2014incorporating,Fellows2020}. For a cross-sectional sample, HIV incidence at the population level may be estimated by using the HIV testing histories and the estimated time between infection; the approach has been applied to the AIDS Indicator Survey (AIS) \citep{Fellows2020}.

As global HIV prevention and treatment programs increase, more data sources are becoming available. Starting from 2015, the U.S. President's Emergency Plan for AIDS Relief (PEPFAR) supported Population-based HIV Impact Assessment (PHIA) individual-level surveys in the most-affected countries in sub-Saharan Africa. PHIA is a nationally-representative HIV-focused survey that combines household visits with key questions and cutting-edge technologies such as biomarker tests for HIV antibody and HIV viral load \citep{MPHIA2018}. These additional biomarker tests offer the unique opportunity of distinguishing between recent infection, i.e. infected within one year, and long-term infection, i.e., infected over one year ago \citep{Hallett2011,WHO2020}, and providing relevant HIV information by age, gender, and location \citep{CDC2017,Hader2016,Justman2018}.

In the PHIA data, however, the time since HIV infection is still unknown. So a classification method is needed to determine the HIV recency status of tested individuals. As its current method of HIV recency prediction, PHIA uses a deterministic binary classification tree (BCT) to assign individuals as ``Recent" or ``Long-term".  A ``Recent" infection has low normalized optical density (ODn $\leq$ 1.5) from HIV-1 limiting antigen avidity enzyme immunoassay, high viral load (VL $\geq$ 1000 copies/mL), and no antiretroviral treatment (ART) detected \citep{MPHIA2018}. Note that the BCT classification can be problematic with small sample sizes because recent infection are rare events among HIV positive individuals. For instance, BCT often produced zero estimates when dis-aggregating HIV incidence by age groups or sub-national areas where no individual is flagged as ``Recent". Therefore, a probabilistic prediction model is desirable when the sample size is relatively small. Unfortunately, the regression model does not work on PHIA data for a simple reason: while it contains biomarker test results which are strong predictors for HIV recency status together with many other individual-level covariates, it does not contain the true recency status for any individual.

To compensate for the unknown HIV recency status in PHIA, we propose a novel statistical model that borrows information from existing cohort studies where the label (recency status) is known at the aggregated level (which we summarize as a contingency table). 
These longitudinal cohort studies provide detailed information on how covariates relate to recency status \citep{Duong2015, Kassanjee2016}. Ideally, these cohort studies would be conducted for each subnational area and subpopulation of interest, but that may be too expensive.
Applying our model to the 2015-2016 Malawi PHIA survey and utilizing the results of existing cohort studies, we obtain the probabilistic individual-level predictions, which is more accurate than BCT predictions and enables better estimation for sub-populations with small sample sizes.

Our classification method falls under the umbrella of weakly supervised learning in which the training data is incomplete, inexact, or inaccurate \citep{Zhou2018}.  In our problem, obtaining labels for each observation is very challenging because people might not know when they were infected or hesitate to share that information.  While we do have some labeled data from the literature, our labeled data by itself would have limitations in fitting a classification model because they are summarized in form of a contingency table.
Learning from set-level labels or summary statistics has been seen in the literature such as multiple-instance learning \citep{Carbonneau2018,Zhou2018}, aggregate output learning \citep{Musicant2007} and Bayesian large-scale multiple regression with summary statistics \citep{Zhu2017}. 
In this article, we decide to utilize both labeled data (existing cohort studies) and unlabeled data (individual-level PHIA) to train our model, and combine the strengths of both. This inductive semi-supervised learning approach \citep{Van2020} has been used in other cases such as the PU (positive-unlabeled) learning \citep{Ward2009, Song2019, Elkan2008, Liu2003}.

As a challenge to combining the labeled and unlabeled data, the covariates are mismatched in both granularity and dimension.  In terms of granularity, the labeled data covariates are discretized, while our unlabeled data covariates are continuous.  In terms of the number of covariates, the labeled data only contain one or two covariates, while the unlabeled data contain a much larger set of covariates (more than one hundred) including those in the labeled data.  We are interested in whether the additional non-discretized covariates in the unlabeled data could improve the prediction accuracy of the labeled data.  To fit a logistic regression on this mismatched data, we assume the datasets share some conditional probability distributions, and modify the logistic regression model.




In the rest of the paper, we describe our model details in Section \ref{sec:Method} and data sources in Section \ref{sec:data}.  Section \ref{sec:RealDataResult} presents the data analysis results for Malawi, and Section \ref{sec:Simulation} uses a simulation study to compare our model performance with a fully supervised learning method.  Finally, we conclude with a discussion in Section \ref{sec:Conclusion}.

\section{Logistic regression without labeled individual-level training data}
\label{sec:Method}

We propose a semi-supervised logistic regression with labeled external summary data, i.e., contingency table, and use that to analyze an unlabeled primary dataset. We provide model details in the following order.  We first define the statistical models for each dataset, including a logistic regression on the primary dataset and a multinomial distribution on the external contingency table.  The primary dataset does not have response labels, and we compensate by combining the primary dataset with a contingency table.  We use maximum likelihood estimation to estimate the model coefficients, and adjust the population proportion $P(Y=1)$, and use information criterion to select the covariates for the logistic regression model.  Finally, we present the asymptotic properties of our proposed estimator.


\subsection{Statistical models for different types of data}
\label{subsection: data models}

In this subsection, we describe the statistical model for each dataset separately.  
We define a standard logistic regression on primary dataset individuals and a multinomial distribution on the contingency table counts.  Both models contain the binary response $Y \in \{0,1\}$ and covariates $X$.  Some covariates $\{X_{1},\cdots,X_{r}\}$ are present in both datasets, while other covariates $\{X_{r+1},\cdots,X_{p}\}$ are only in the primary dataset.

\subsubsection{Logistic regression model for the primary dataset}
\label{subsubsection:logistic}

The primary dataset consists of the covariates $X_{i.}=(X_{i1},\cdots,X_{ip})^T$ for individuals $i=1,\cdots,n$. The individual responses $Y_i$ are of interest, but completely unobserved in the primary dataset.  We define the following logistic regression to model the probability of $Y_i=1$:
\begin{equation}
    P(Y_i=1|X_{i.};\beta) = logit^{-1}(\beta_0 + X_{i1} \beta_1 + \cdots + X_{ip} \beta_p), \; \; i=1,\cdots,n,
\label{eqn:logistic1}
\end{equation}
where $\beta$'s are the regression coefficients.  The intercept $\beta_0$ is always included, and  non-intercept regression coefficients can be either nonzero or zero. 

 
\subsubsection{Model for contingency table}
\label{subsubsection:multinomial}

Existing studies reveal the relationship between $Y$ and some discretized covariates in the form of a contingency table.  The contingency table is defined by table covariates $T=(X_1, \cdots, X_r)$, where $r \le p$, and these table covariates are discretized into $K$ disjoint table cells $C_1, \cdots, C_K$.  Each cell $C_k$ is a subset of the table covariate space, and the union of cells covers the entire $T$ space.  Each table cell $C_k$, $k=1,\cdots,K$, contains observed counts $m^{(1)}_k$ and $m^{(0)}_k$ for $Y=1$ and $Y=0$.


We define a multinomial distribution on the contingency table counts $m^{(h)}_k$, and show the multinomial likelihood $L$ (without the multinomial coefficient):
\begin{equation}
    L = \prod_{h=0}^1 \prod_{k=1}^K P(Y=h, T \in C_k)^{m^{(h)}_k},
   \label{eqn:Multinomial}
\end{equation}
where $m^{(h)}_k$ is the count corresponding to $Y=h$ and $T \in C_k$.

\subsection{Proposed model}
\label{subsection: combine data}

The individual responses $Y_i$'s are completely unobserved in the primary dataset, so we could not estimate $\beta$'s using only the primary dataset.  To overcome this challenge, we incorporate information from the contingency table, which contains observed values of $Y$.  This enables us to estimate $\beta$, and predict individual $Y_i$ in the primary dataset.

\subsubsection{Model assumptions and likelihood}
\label{subsubsection: combined likelihood}

In this subsection, we introduce additional assumptions that allow us utilize information from both datasets. We first assume that the relationship between X and Y is the same between the two datasets.  That is, for both datasets, we assume that the conditional distribution of $Y|X$ follows Equation (\ref{eqn:logistic1}), except for the intercept $\beta_0$ which will be adjusted post-estimation in Section \ref{subsubsection: beta estimation}.

In the contingency table, the logistic regression covariates $\{X_{1},\cdots,X_{p}\}$ are either in discrete form $\{T \in C_k\}$ or unavailable, and the covariates mismatch with the primary dataset in both granularity and dimension.  To resolve this mismatch, we further assume the contingency table and the primary dataset have the same $P(X|T \in C_k)$.  Note that we did not assume the same marginal distribution of covariates $X$ (or same marginal $P(T)$) between data sources but the same conditional distribution given $T \in C_k$, and this assumption is reasonable when the range of $C_k$ is small.  In Equation (\ref{eqn:empirical X}), we approximate $P(X|T \in C_k)$ by its empirical distribution in the primary dataset.  

\begin{equation}
P(X|T \in C_k) \approx \frac{w_i}{\sum_{i: T_{i} \in C_k} w_i},
\label{eqn:empirical X}
\end{equation}
where $T_{i}=\{X_{i1},\cdots,X_{ir}\}$, and $w_i$ is the sampling weight of individual $i$ such that $\sum_{i=1}^n w_i = 1$.

Combining our assumptions, the primary dataset and contingency table have the same $P(Y|T \in C_k; \beta)$.  (At this stage, we treat the two datasets as having the same intercept $\beta_0$, but this will be adjusted post-estimation in Section \ref{subsubsection: beta estimation}.)  We approximate $P(Y|T \in C_k; \beta)$ from the primary dataset, using Equations (\ref{eqn:logistic1}) and (\ref{eqn:empirical X}):
\begin{equation}
   P(Y=h|T \in C_k; \beta) \approx \sum_{i: T_{i} \in C_k}  \frac{P(Y_i=h|X_{i.};\beta)\times w_i}{\sum_{i: T_{i} \in C_k} w_i}.
   \label{eqn:approximation of p_k}
\end{equation}

Finally, we combine the logistic regression with the contingency table counts $m^{(h)}_k$ in our likelihood from Equation (\ref{eqn:Multinomial}).  We remove the marginal $P(T \in C_k)$ from our likelihood because we match the two datasets based on same conditional $P(Y|T \in C_k; \beta)$, not same joint $P(Y,T \in C_k; \beta)$.  (The two datasets may have different marginal $P(T \in C_k)$.)  This will not affect estimation of $\beta$, since the marginal distribution of $T$ does not depend on $\beta$.  We show the conditional multinomial likelihood:  
\begin{equation}
    L(\beta) = \prod_{k=1}^K \left( P(Y=1|T \in C_k; \beta)^{m^{(1)}_k} \times P(Y=0|T \in C_k; \beta)^{m^{(0)}_k} \right),
   \label{eqn: Conditional Multinomial}
\end{equation}
where $m^{(h)}_k$ is the count corresponding to $Y=h$ and $T \in C_k$.  In the likelihood, we substitute $P(Y=h|T \in C_k; \beta)$ with the approximation from Equation (\ref{eqn:approximation of p_k}), so the likelihood now includes both datasets.

\subsubsection{Coefficient estimation}
\label{subsubsection: beta estimation}


We first estimate $\beta$'s by maximizing the approximate conditional multinomial likelihood from Equations (\ref{eqn:approximation of p_k}) and (\ref{eqn: Conditional Multinomial}).  To maximize the likelihood, we use the $optim()$ numerical optimization function in $R$, and use the Nelder-Mead method within $optim()$.  The $optim()$ function also provides a ``numerically differentiated Hessian", and we invert the negative Hessian to obtain the covariance matrix, from which we compute the standard error and correlation of coefficients \citep{R2020}. The resulting maximum likelihood estimate (MLE) is $\hat{\beta} = \arg\max L(\beta)$ for the contingency table data.

The primary dataset is representative of the population that we would like to make inference, but the contingency table data may oversample $Y=1$.  To make proper prediction for the primary data, the intercept $\beta_0$ shall be calibrated, and this follows standard case-control adjustment procedure.  In epidemiology, the population-level estimator of prevalence or incidence is sometimes available, and can be used to calibrate the intercept of the case-control study \citep{Breslow1996}. Let $\hat{P}(Y=1)$ be such an estimator. We re-estimate $\hat{\beta_0}$ such that 
\begin{equation}
\label{eqn:case_control_adjustment}
    \sum_{i=1}^n P(Y_i=1|X_{i.}; \hat{\beta}) \times w_i = \hat{P}(Y=1),
\end{equation}
where non-intercept $\hat{\beta}$ are fixed at MLE. With the calibrated $\hat{\beta}_0$, we can calculate the estimated  probabilities of $Y=1$ for the primary data, $P(Y_i=1|X_i;\hat{\beta})$.

\subsubsection{Consistency and Central Limit Theorem}

In this section, we establish the theoretical properties of our estimator. Let $m = m_0+m_1$ be the number of total observations in the contingency table, where $m_0$ and $m_1$ are the numbers of observations with $Y=0$ and $1$, respectively.
Denote the true probability in cell $(T\in C_k,Y=h)$ of the contingency table by $p^{(h)}_k({\beta}^*)$, where $\beta^*$ are the true logistic regression intercept for the contingency table.
More specifically,
we have
\begin{equation}
p^{(h)}_k({\beta}) = P(T \in C_k , Y=h) 
= \int_{C_k} P(Y=h | {X} ) \, d \mu_X 
= \int_{C_k} \frac{ 1 - h + h \exp({X}^T {\beta}) }
{1 + \exp({X}^T {\beta})}  \, d \mu_X ,
\end{equation}
where $\mu_X$ is the support of $X$.
We impose the following conditions for the consistency of our proposed estimator:
\begin{description}


\item[A1]
$P(T \in C_k, Y=h) \neq 0$ for any $h=0,1$ and $k= 1,\cdots, K$.

\item[A2] There is no $\beta \neq {\beta}^*$ such that 
$p^{(h)}_k({\beta}) = p^{(h)}_k({\beta}^*)$ for $h=0,1$ and $k=1,\cdots,K$.

\item[A3] 
All predictors $X_1, \cdots, X_p$ have finite second moments.


\end{description}

\noindent\textbf{Remark}:
Conditions~A1 and A2 are for the identifiability of our model. They imply that $2K-1 \geq p$, where $2K$ is the number of the contingency table cells and $p$ is the dimension of $\beta$. That is to say, the contingency table should provide enough degrees of freedom ($2K-1 \geq p$) for estimating $p$ logistic regression coefficients.
Conditions~A3 is a quite mild regularity condition for the distribution of ${X}$. 

\begin{theorem}
\label{theorem:consistency}

Under Conditions~A1, A2, and A3,
then the estimator $\hat{{\beta}}$  converges to the true logistic regression coefficient vector ${\beta}^*$ in probability as $m \rightarrow \infty$ and $n \rightarrow \infty$.
Furthermore, we have the following asymptotic distribution for $\hat{{\beta}}$:
\begin{equation}
	\hat{{\beta}}-{\beta}^* = 
	{H} ({R}_Z^T, {R}_W^T)^T, 
\end{equation}
and 
\begin{equation}
\begin{aligned}
	\left(
	\sqrt{n} {R}_Z^T,
	\sqrt{m} {R}_W^{T}
	\right)^T
	\overset{d}{\rightarrow} 
	N({0}, 
	\Diag({V}_Z, {V}_W)) ,
\end{aligned}
\end{equation}
as $n, m, \rightarrow \infty$,
where 
${H}$, ${V}_Z$, ${V}_W$ are constant matrices of dimension $p\times (6K)$, $(4K)\times(4K)$, and $(2K)\times(2K)$, which only depend on the distribution $\mu_X$, the true logistic regression coefficient ${\beta}^*$, 
and the cells $C_k$,
$R_Z$ and $R_W$ are two random vectors of lengths $4K$ and $2K$, respectively.
The proof is provided in Appendix A1.
\end{theorem}

Theorem~\ref{theorem:consistency} establishes the consistency and central limit theorem for $\hat{\beta}$.
It shows that the difference of $\hat{\beta}$ and the true coefficient vector can be decomposed into two parts $R_Z$ and $R_W$, which measure the contribution of randomness from the primary data set and the contingency table, respectively.

After the estimation of $\beta^{*}$, we use case-control adjustment \eqref{eqn:case_control_adjustment} to estimate $\beta_0^\textup{prim}$, where $\beta_0^\textup{prim}$ is the intercept of the logistic regression coefficient on the primary data set.
Theorem~\ref{theorem:case_control_adjustment} further quantifies the order of difference between the case-control adjustment estimator of the logistic regression intercept and its true value.

\begin{theorem}
\label{theorem:case_control_adjustment}
Under Conditions~A1, A2, and A3, we have
\begin{equation}
\hat{\beta}_0 - \beta_0^\textup{prim}
=O_p\left(\sqrt{\frac{1}{n}+\frac{1}{m}+(\hat{P}(Y=0)-P(Y=0))^2} \right) ,
\end{equation}
where $\hat{\beta}_0$ is the case-control adjustment estimator from \eqref{eqn:case_control_adjustment} and $\beta_0^\textup{prim}$ is the true intercept for the logistic regression on the primary data set.
The proof is provided in Appendix A2.
\end{theorem}

\subsection{Model selection}

To select the covariates in the logistic model, we perform forward stepwise AIC (Akaike Information Criterion) or BIC (Bayesian Information Criterion) starting from the intercept-only model \citep{Dziak2020}.  At each step, out of the non-selected covariates, we select the covariate that would decrease AIC (or BIC) the most, and add it to the model.  We repeat this step until no covariate would decrease AIC (or BIC) further.

When the covariates include missing values, we address the uncertainty of the imputation in both the model selection and the parameter estimates. Denote $J$ as the number of imputed datasets. For each imputed dataset $j=1,\ldots,J$, we run forward stepwise AIC/BIC, and select the best set of covariates $A_j$. Then, we apply the majority criteria and keep the covariates that are selected in at least half of the imputed datasets \citep{Van2018}.  Collectively, these kept covariates form $A_{final}$, the set of covariates in the final consensus model. With $A_{final}$, we refit the coefficients $\hat{\beta}$ for each imputed dataset, and compute the corresponding individual probabilities $P(Y_i=1|X_{i.};\hat{\beta})$ for each imputed dataset. To summarize $\hat{\beta}$ and $P(Y_i=1|X_{i.};\hat{\beta})$, we average their values over the $J$ imputed datasets. To compute the standard errors of $\hat{\beta}$, we assume that the coefficient estimates have multivariate normal distributions and apply Rubin's rules to compute total variance \citep{Rubin1987}.

\section{Data sources}
\label{sec:data}

We introduce the primary dataset \,---\, 2015-2016 Malawi PHIA in Section \ref{sec:primary-data}.  In Section \ref{sec:external-data}, we discuss the contingency tables that summarize the relationship between HIV recency $Y$ and some covariates, where $Y=1$ represents a true recent infection (infected within one year) and $Y=0$ represents a true long-term infection.

In addition, we obtain the 2016 Malawi HIV incidence estimate (4.38 per 1000 susceptible) and HIV prevalence estimate (9.9\%) from the UNAIDS website \citep{UNAIDS2021}.  From these two estimates, we obtain the proportion of HIV positive individuals who are recently infected. Individuals who receive antiretroviral treatment (ART) are generally considered as being infected for at least one year; many newly infected individuals do not get tested and diagnosed until months or years after infection \citep{Jones2019, Giguere2021}, and despite guidelines increasing ART eligibility, there may still be a gap between HIV diagnosis and treatment initiation \citep{Esber2020}. Excluding those on ART, we obtain an estimate of the marginal probability, $\hat{P}(Y=1)=0.126$, which is the probability of being recently infected among those who are HIV positive but not on ART. In Appendix B, we show a more detailed derivation of $\hat{P}(Y=1)$.

\subsection{Primary dataset: Malawi PHIA}
\label{sec:primary-data}

The primary dataset, 2015-2016 Malawi PHIA, is a cross-sectional individual-level survey for evaluating Malawi's national HIV strategy, in part by estimating HIV incidence.  The survey is based on a two-stage, stratified cluster sample design with individual sampling weights. PHIA is nationally representative for people living with HIV in Malawi \citep{MPHIA2018, Mtech2018}.

We focus on the adult population (ages 15-64) for HIV incidence estimation \citep{MPHIA2018}, and subset to $n=705$ HIV positive individuals with undetectable ART.
The subsetting matches both the binary classification tree described in Section \ref{sec:Intro} and the contingency tables \citep{MPHIA2018, Duong2015, Kassanjee2016}. 
Following the Malawi PHIA report, we use individual blood sampling weights in the Adult Biomarker dataset. These weights account for sample selection probabilities, and are adjusted for noncoverage and nonresponse \citep{MPHIA2018, Mtech2018}.  

While PHIA does not have the binary response $Y$ of recent infection status, the data does contain individual-level covariates.  We consider the following PHIA covariates: ODn (normalized optical density) from HIV-1 limiting antigen avidity enzyme immunoassay, VL (viral load), CD4 count, age, gender, self-reported HIV status, and HIV testing status.  
\begin{itemize}
    \item ODn, VL, CD4, and age are continuous covariates.  We take the logarithm of VL and the 4th root of CD4 counts as suggested in \citep{Pantazis2019}. 659 individuals have numeric VL values. For non-numeric VL, we set ``undetectable" to 20 for 39 individuals, and set ``less than 40" to 30 for 7 individuals. 
    \item The following variables are binary: gender, self-reported HIV positive, self-reported HIV negative, tested for HIV (and received positive results) within 1 year, and the interaction of self-reported HIV negative and tested for HIV within 1 year.
\end{itemize}
We use multiple imputation to fill in missing covariate values \citep{Van2018}. For CD4, 24 individuals have missing values (681 have complete CD4 values), and we perform multiple imputation 100 times for the 24 missing values and thus create 100 datasets with imputed CD4 values. Additionally, we keep the geographic variables of region and zone for aggregating the HIV incidence estimates at different sub-national levels. 

In Appendix C, we show a more detailed processing of the primary dataset.

\subsection{External dataset: contingency table}
\label{sec:external-data}

Our external datasets are contingency tables that show the number of specimens in covariate groups given true HIV recency status. The table data are based on cohort studies with blood specimens drawn at different times post-infection \citep{Duong2015, Kassanjee2016}.

We use two contingency tables: a univariate contingency table based on ODn and a bivariate contingency table based on (ODn,VL).  The contingency table counts $m^{(h)}_k$, $k=1, \cdots, K$, are mostly non-integer because they are not directly observed but derived from the mean duration of recent infection and the false recency rate. We show the derivation in Appendix D.

We derive the univariate ODn contingency table from Table 2 of Duong 2015 \citep{Duong2015}. The sample size of $Y=1$ is 994, and the sample size of $Y=0$ is 3739.  For this table, the $Y=1$ and $Y=0$ specimens come from different studies. We show the univariate ODn contingency table in Table \ref{tab:uniCT}.  

\begin{table}[h]
\centering
\renewcommand{\thetable}{1.1}
\begin{tabular}{|c|c|c|} 
\hline
ODn $(\gamma_1,\gamma_2)$ & $Y=1, \gamma_1 < ODn \leq \gamma_2$ & $Y=0, \gamma_1 < ODn \leq \gamma_2$ \\ \hline
\textbf{(0,1]} & 239.6 & 24 \\ \hline
\textbf{(1,1.25]} & 57.2 & 13 \\ \hline
\textbf{(1.25,1.5]} & 57.2 & 23 \\ \hline
\textbf{(1.5,1.75]} & 40.8 & 10 \\ \hline
\textbf{(1.75,2]} & 43.6 & 23 \\ \hline
\textbf{(2,$\infty$)} & 555.6 & 3646 \\ \hline
\end{tabular}
\caption{Univariate ODn contingency table. The cells provide the counts of observations that belong to the corresponding categories.}
\label{tab:uniCT}
\end{table}

We derive the bivariate (ODn,VL) contingency table from Supplementary Digital Content 2 of Kassanjee 2016 \citep{Kassanjee2016}. The sample size of the $Y=1$ is 627, and the sample size of $Y=0$ is 873.  For this table, the $Y=1$ and $Y=0$ specimens come from the same study.  However, this study draws relatively more specimens at the beginning vs the end, and oversample $Y=1$.  We show the bivariate (ODn,VL) contingency table in Table \ref{tab:biCT}.  

\begin{table}[!h]
\centering
\renewcommand{\thetable}{1.2}
\begin{tabular}{|c|c|c|c|c|}
\hline
\multicolumn{2}{|c|}{$Y=1, \kappa_1 \leq VL < \kappa_2,$} & \multicolumn{3}{c|}{VL ($\kappa_1,\kappa_2$)} \\ \cline{3-5}
\multicolumn{2}{|c|}{$\gamma_1 < ODn \leq \gamma_2$} & \textbf{{[}0,1000)} & \textbf{{[}1000,$\infty$)} & \textbf{{[}0,$\infty$)} \\ \hline
\multirow{8}{*}{ODn ($\gamma_1,\gamma_2$)} & \textbf{(0,0.5{]}} & 25.8 & 75.6 & NA \\ \cline{2-5} 
& \textbf{(0.5,1{]}} & 22.3 & 68.7 & NA \\ \cline{2-5} 
& \textbf{(1,1.5{]}} & 6.9 & 73.9 & NA \\ \cline{2-5} 
& \textbf{(1.5,2{]}} & 3.4 & 36.1 & NA \\ \cline{2-5} 
& \textbf{(2,2.5{]}} & 6.9 & 80.7 & NA \\ \cline{2-5} 
& \textbf{(2.5,3{]}} & 1.7 & 77.3 & NA \\ \cline{2-5} 
& \textbf{(3,4.5{]}} & 6.9 & 108.2 & NA \\ \cline{2-5} 
& \textbf{(4.5,$\infty$)} & NA & NA & 32.6 \\ \hline
\hline
\multicolumn{2}{|c|}{$Y=0, \kappa_1 \leq VL < \kappa_2,$} & \multicolumn{3}{c|}{VL ($\kappa_1,\kappa_2$)} \\ \cline{3-5}
\multicolumn{2}{|c|}{$\gamma_1 < ODn \leq \gamma_2$} & \textbf{{[}0,1000)} & \textbf{{[}1000,$\infty$)} & \textbf{{[}0,$\infty$)} \\ \hline
\multirow{8}{*}{ODn ($\gamma_1,\gamma_2$)} & \textbf{(0,0.5{]}} & 4.4 & 2.6 & NA \\ \cline{2-5} 
& \textbf{(0.5,1{]}} & 0.9 & 7.0 & NA \\ \cline{2-5}
& \textbf{(1,1.5{]}} & 9.6 & 19.2 & NA \\ \cline{2-5}
& \textbf{(1.5,2{]}} & 6.1 & 17.5 & NA \\ \cline{2-5}
& \textbf{(2,2.5{]}} & 9.6 & 47.1 & NA \\ \cline{2-5}
& \textbf{(2.5,3{]}} & 5.2 & 91.7 & NA \\ \cline{2-5}
& \textbf{(3,4.5{]}} & 22.7 & 340.5 & NA \\ \cline{2-5}
& \textbf{(4.5,$\infty$)} & NA & NA & 289.0 \\ \hline
\end{tabular}
\caption{Bivariate (ODn,VL) contingency table. The cells provide the counts of observations that belong to the corresponding categories.}
\label{tab:biCT}
\end{table}


\section{Estimating HIV incidence from Malawi PHIA data}
\label{sec:RealDataResult}

We apply our proposed method to datasets presented in the previous section. They are contingency tables, both univariate ODn of Duong 2015 \citep{Duong2015} and bivariate (ODn,VL) of Kassanjee 2016 \citep{Kassanjee2016}, 2015-2016 Malawi PHIA \citep{MPHIA2018}, and the estimated $\hat{P}(Y=1)=0.126$.

In Section \ref{sec:results selection}, we present the variable selection results when only using the univariate ODn table, when only using the bivariate (ODn,VL) table, and when using both tables. With those results, we illustrate that more covariates could be included in the logistic regression model besides the table covariates.  In Section \ref{sec:results manual}, we build our classifier using both tables, and show that our classifier outperforms the BCT, at both individual and aggregated levels.


\subsection{Variable selection}
\label{sec:results selection}

We run our method on the real contingency tables, and use forward stepwise AIC/BIC to select the model covariates for the logistic model.  We fit the univariate ODn contingency table, the bivariate (ODn,VL) contingency table, and combined univariate and bivariate tables.

\subsubsection{Results of univariate contingency table}

We fit the proposed model to the univariate ODn contingency table data which is simpler and easier to understand. It contains a relatively larger sample size (4733 specimens) than the bivariate (ODn,VL) contingency table data (1500 specimens). 

AIC selects a larger set of predictors than BIC.  Starting with the intercept, we add covariates to the model.  Using AIC, for all 100 imputed datasets, we select two covariates ODn and tested for HIV within 1 year, and then stop. Using BIC, for all 100 imputed datasets, we select one covariate, ODn, and then stop. The AIC values progress from 4867.3 (intercept only) to 3750.6 (add ODn) to 3748.7 (add tested for HIV within 1 year), while the BIC values progress from 4873.8 (intercept only) to 3763.5 (add ODn).  As a significant note, our stepwise AIC procedure selects the covariate tested for HIV within 1 year, even though that covariate is not part of the univariate ODn contingency table; this shows that our method can select auxiliary covariates that are useful in predicting the HIV recency status but not available in the external contingency table.

Using the model selected by AIC (intercept, ODn, and tested for HIV within 1 year), we obtain the fitted approximate probabilities $P(Y=h|T \in C_k; \beta)$ based on the estimated coefficients $\hat{\beta}$, and these are the probabilities used in the multinomial likelihood.  In Table \ref{tab:fitUniCPT}, we compare the fitted distributions $P(Y=h|T \in C_k; \beta)$ to the data distributions $\frac{m^{(h)}_k}{m^{(0)}_k+m^{(1)}_k}$ from Duong 2015, and find that the distributions are similar, especially for cells where the cell count, $m^{(h)}_k$, is large \citep{Duong2015}.

\begin{table}[!h]
\centering
\renewcommand{\thetable}{2}
\begin{tabular}{|c|c|c|c|c|}
\hline
\multirow{2}{*}{$(\gamma_1,\gamma_2)$} & \multicolumn{2}{c|}{$P(Y=1| \gamma_1 < ODn \leq \gamma_2)$} & \multicolumn{2}{c|}{$P(Y=0| \gamma_1 < ODn \leq \gamma_2)$} \\ \cline{2-5} 
 & Data & Fitted & Data & Fitted \\ \hline
\textbf{(0,1]} & .909 & .909 & .091 & .091 \\ \hline
\textbf{(1,1.25]} & .815 & .823 & .185 & .177 \\ \hline
\textbf{(1.25,1.5]} & .713 & .716 & .287 & .284 \\ \hline
\textbf{(1.5,1.75]} & .803 & .802 & .197 & .198 \\ \hline
\textbf{(1.75,2]} & .655 & .641 & .345 & .359 \\ \hline
\textbf{(2,$\infty$)} & .132 & .132 & .868 & .868 \\ \hline
\end{tabular}
\caption{Comparison between the empirical proportions (``Data") and the fitted proportions (``Fitted") for univariate ODn contingency table}
\label{tab:fitUniCPT}
\end{table}

\subsubsection{Results of bivariate contingency table}

The bivariate (ODn,VL) contingency table contains one more table covariate than the univariate ODn contingency table. Using AIC, for 93 out of 100 imputed datasets, we select three covariates, ODn, VL and self-reported HIV negative.  The other 7 imputed datasets include CD4 as a selected covariate: 2 of them select ODn, CD4, self-reported HIV negative; 5 of them select ODn, CD4, self-reported HIV negative, and the interaction between self-reported HIV negative and tested for HIV within 1 year. Using BIC, for all the 100 imputed datasets, we select one covariate, ODn. The AIC values progress from 2040.9 (intercept only) to 1504.3 (add ODn) to 1500.4 (add VL) to 1494.6 (add self-reported HIV negative), while the BIC values progress from 2046.2 (intercept only) to 1514.9 (add ODn).  Although the bivariate (ODn,VL) contingency table contains fewer subjects, less than one third of the univariate ODn contingency table, it could possibly support a model with higher degrees of freedoms, based on AIC.

\subsubsection{Results of combined tables}

Finally, we use all the available data, and fit our model on both contingency tables. The two cohort studies summarized by the univariate and the bivariate tables are independent. So the parameters are estimated by maximizing the product of univariate table likelihood and the bivariate table likelihood.

Using AIC, for 97 out of 100 imputed datasets, we select three covariates: ODn, VL, and self-reported HIV positive. The other 3 imputed datasets select four covariates: ODn, then VL, then self-reported HIV positive, and finally, CD4. Using BIC, for all the 100 imputed datasets, we select one covariate, ODn.  The AIC values progress from 7146.6 (intercept only) to 5268.0 (add ODn) to 5263.9 (add VL) to 5261.9 (add self-reported HIV positive), while the BIC values progress from 7153.3 (intercept only) to 5281.5 (add ODn).

\subsubsection{Summary of variable selection}
In the above analyses, BIC selects fewer covariates than AIC because its penalty depends on the sample size.  BIC always selects only one covariate, ODn.  
Under the BIC selection, our method reduces to something similar to the traditional recency assay analysis with the false positive / negative rates determined by ODn cutoffs, except we use the ODn covariate as a continuous predictor.
For the bivariate and combined table analyses, BIC misses the table covariate VL, which is known to be associated with HIV recency status. We think the BIC penalty term, the number of parameters $\times \log(\textup{sample size of contingency table})$, might be too strong when the response are labeled for groups of individuals whose covariates are in the same range.

AIC has suggested additional covariates besides the table covariates: tested for HIV within 1 year for the univariate contingency table, self-reported HIV negative for the bivariate contingency table, and self-reported HIV positive for combined tables. The first two are indicators of relatively recent HIV infection while the last one suggests a relatively long-term HIV infection. CD4 is also selected among a few imputed datasets (7\% for the bivariate table and 3\% for the combined tables), and it is known to be an important indicator of HIV/AIDS progression \citep{Cori2015}. We suspect that the predictive power of CD4 is somewhat limited by its missingness. 


\subsection{Result Comparison with BCT}
\label{sec:results manual}



In this subsection, we compare the results between our proposed logistic regression and the binary classification tree (BCT) at both individual and area levels. For the logistic regression, we use both the univariate table and the bivariate table as our labeled training data. The BCT classifies a non-ART HIV positive individual as ``Recent" infection if ODn $\leq$ 1.5 and VL $\geq$ 1000, and as ``Long-Term" otherwise \citep{MPHIA2018}. We validate the individual predictions in MPHIA data in \ref{sec:individual}, and present the estimated HIV recency rates among different groups in \ref{sec:aggregated}.

\subsubsection{Fitted individual-level HIV recency}
\label{sec:individual}

We use ODn and VL as the logistic regression covariates so that both the logistic regression and the BCT are defined on the same set of covariates. Fitting our logistic regression on (ODn,VL), we obtain coefficient estimates $(\hat{\beta_0},\hat{\beta_{ODn}},\hat{\beta_{VL}})=(-2.82, -1.56, .21)$ ($\hat{\beta_0}$ adjusted from -1.72) with SE $(SE(\hat{\beta_0}), SE(\hat{\beta_{ODn}}),SE(\hat{\beta_{VL}})) = (.05, .06, .08)$.  We generally expect a recently infected individual to have low ODn and high VL, and both classifiers predict with this expectation.  For each individual in our Malawi PHIA dataset, the logistic regression model provides the fitted probability of having a recent HIV infection, which can be compared with the classification results provided by BCT.



\begin{figure}[!h]
\centering
    \includegraphics[width=.85\textwidth]{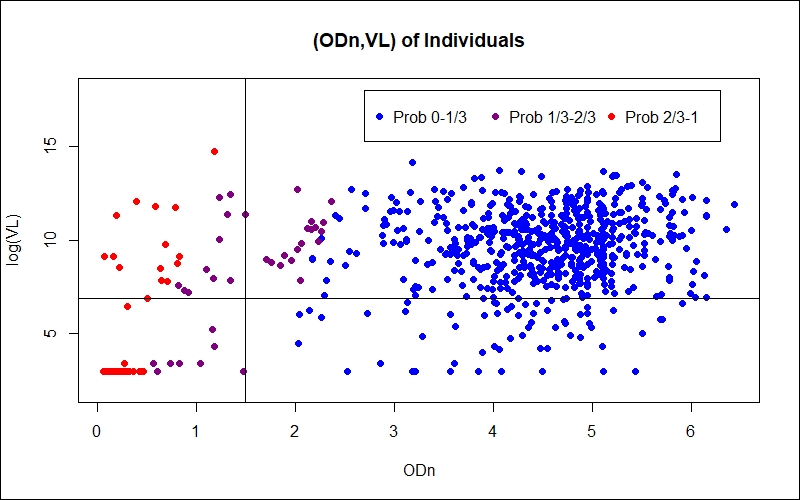}
    \caption{Scatterplot of (ODn,VL) values for the 705 Malawi PHIA individuals.  The three point colors represent the fitted recency probability from our method.  The top left quadrant represents BCT ``Recent", while the other three quadrants represent BCT ``Long-Term".}
    \label{fig:scatterplot}
\end{figure}

In Figure \ref{fig:scatterplot}, we plot the (ODn,VL) values of the 705 Malawi PHIA individuals as scatterplot points.  For the BCT predictions, we divide the graph into four quadrants with lines ODn=1.5 and VL=1000; the top left quadrant is BCT ``Recent", and the other three quadrants are BCT ``Long-Term".  For our method predictions, we color the points based on the fitted individual probabilities: blue for probabilities $0\sim 1/3$, purple for probabilities $1/3\sim2/3$, and red for probabilities $2/3\sim 1$.  As expected from our logistic model, fitted probability increases from bottom right to top left. 11 BCT ``Recent" individuals seem to have lower fitted logistic probability (purple), but upon closer investigation, the 11 probabilities range from .53 to .66, which is still high enough to suggest recent infection. Some BCT ``Long-Term" individuals have high recency probability prediction; out of 680 individuals, 25 have fitted probability between $1/3\sim2/3$, and 31 have fitted probability between $2/3\sim 1$.  For many of these individuals, their (ODn,VL) points lie just outside the boundary of the BCT ``Recent" box.  Although BCT requires VL$\ge 1000$ in order to label a recent infection, our model suggests that such requirement may not be needed for individuals with ODn$<0.5$. Those individuals with low VL actually could be recently infected because the VL rapidly decreases 3-4 weeks after initial infection (as the body develops an immune response) \citep{Lavreys2002, Kahn1998}.


Note that self-reporting HIV negative is a strong indicator of a recent infection because the individual is not yet aware of his/her HIV positive status upon participating in the PHIA survey. In the above analysis, self-reporting HIV negative is not used as a covariate, but instead could serve as a validation indicator. Table \ref{tab:compSRneg} presents the proportions of self-reporting negative stratified by the BCT classifications and by the logistic regression fitted probabilities.
BCT ``Recent" individuals have higher self-reporting negative proportion (72.0\%) than BCT ``Long-Term" individuals (47.8\%). The self-reporting negative proportion also increases with the logistic regression fitted recency probability. Interestingly, among BCT ``Long-Term" individuals, the self-reporting negative proportions are 68.0\% and 71.0\% for those with fitted probabilities $1/3 \sim 2/3$ and $2/3 \sim 1$ respectively. Those two proportions are much higher than the proportion among those with fitted probability $0 \sim 1/3$ (45.8\%), and fairly close to the proportion among BCT ``Recent" individuals (72.0\%). Therefore, we believe many individuals who are classified as "Long-Term" by BCT but with high logistic regression fitted recency probabilities are actually recently infected.

On the other hand, self-reporting HIV positive suggests a long-term infection because the individual is already aware of his/her HIV positive status.  (To clarify, ``HIV positive" and ``HIV negative" are not the only two self-reporting possibilities.  An individual could report a third category of never tested or never received test results, and a few individuals are missing information on self-reporting.)  BCT ``Long-Term" individuals have a higher self-reporting positive proportion (30.7\%) than BCT ``Recent" individuals (8.0\%), and the self-reporting positive proportion also decreases with the logistic regression fitted recency probability.  Among BCT ``Long-Term" individuals, the self-reporting positive proportion is 32.9\% for those with fitted probability $0 \sim 1/3$, and is much higher than the proportion among those with fitted probabilities $1/3 \sim 2/3$ (16.0\%) and $2/3 \sim 1$ (0\%).  Therefore, we believe many individuals who are classified as ``Long-Term" by BCT but with low logistic regression fitted recency probabilities are actually long-term infected.

\begin{table}[!h]
\centering
\renewcommand{\thetable}{3}
\begin{tabular}{|c|c|c|c|c|c|c|c|c|c|}
\hline
\multirow{2}{*}{Fitted Probability} & \multicolumn{3}{c|}{All Individuals} & \multicolumn{3}{c|}{BCT ``Recent"} & \multicolumn{3}{c|}{BCT ``Long-Term"} \\ \cline{2-10} 
 & Total & SRpos & SRneg & Total & SRpos & SRneg & Total & SRpos & SRneg \\ \hline
Blue: $0 \sim 1/3$ & 624 & 32.9\% & 45.8\% & 0 & NA & NA & 624 & 32.9\% & 45.8\% \\ \hline
Purple: $1/3 \sim 2/3$ & 36 & 13.9\% & 69.4\% & 11 & 9.1\% & 72.7\% & 25 & 16.0\% & 68.0\% \\ \hline
Red: $2/3 \sim 1$ & 45 & 2.2\% & 71.1\% & 14 & 7.1\% & 71.4\% & 31 & 0\% & 71.0\% \\ \hline
\end{tabular}
\caption{Number of individuals (``Total") and proportions (unweighted) self-reporting HIV positive (``SRpos") and negative (``SRneg") in Malawi PHIA.  We stratify these counts and proportions by fitted recency probability from our method (``Fitted Probability") and by BCT classification (BCT ``Recent" vs BCT ``Long-Term").  ``All Individuals" includes both BCT ``Recent" and BCT ``Long-Term".}
\label{tab:compSRneg}
\end{table}

\subsubsection{Estimated HIV recency rates}
\label{sec:aggregated}

In the previous section, the individual-level probabilities are used to compare our proposed logistic regression model with BCT, so we use the same set of covariates for the logistic regression. Here, we compare area-level estimates, rather than individual-level estimates, so we are less concerned about matching covariates from BCT for comparison.

We take the model selected by AIC as our final model which includes three covariates: ODn, VL, and self-reported HIV positive.  Fitting our logistic regression, we obtain coefficient estimates $(\hat{\beta_0}, \hat{\beta_{ODn}}, \hat{\beta_{VL}}, \hat{\beta_{SRpos}}) = (-2.59,-1.53,0.25,-1.19)$ ($\hat{\beta_0}$ adjusted from -1.45) with SE $(SE(\hat{\beta_0}), SE(\hat{\beta_{ODn}}), SE(\hat{\beta_{VL}}), SE(\hat{\beta_{SRpos}})) = (0.11, 0.06, 0.09, 0.58)$. They suggest that HIV recency status among people living with HIV is positively associated with VL and negatively associated with ODn and self-reported HIV positive.

We summarize the individual-level probabilities by taking the weighted average among individuals within a group. This gives the aggregated proportion of HIV infections that are recent, and the proportions are stratified by age-sex groups and gender-region groups. For each group, we present the sample size, our weighted classifier (mean estimate), and the raw weighted BCT (mean estimate) \citep{MPHIA2018}. (For the BCT to be useful at area level, we must adjust the raw BCT estimates by multiplying them \citep{Duong2015, MPHIA2018}.  However, in this paper, we do not make this adjustment because that is outside the scope of our work.)

We compute the 95\% confidence intervals (C.I.) based on Bootstrap quantiles \citep{Efron1993}. To calculate this, we first uniformly resample PHIA observations with replacement.  We resample 1000 PHIA datasets, and each resampled dataset has same number of observations as the original PHIA.  Note that we use individual sampling weights in coefficient estimation and area-level aggregation, but not in resampling PHIA observations.  Next, for each resampled PHIA dataset, we apply our new method.  We check whether there are PHIA individuals in each contingency table cell, and if not, we decide to combine adjacent table cells.  We estimate the model coefficients, individual probabilities, and area-level point estimates.  Finally, for each area level, we compute the 2.5\% and 97.5\% quantiles of the 1000 point estimates.

In Table \ref{tab:compAgeSex}, we present the demographic summary of age-sex. 
We notice that for certain age-sex groups, no individual is flagged as ``Recent" by BCT, e.g. male 45-54.  As a result, the BCT estimates the HIV recency rates to be zero with zero standard error. Some nonzero BCT estimates are also surprisingly low due to the very small number of ``Recent" individuals, e.g. female 35-44 has an estimated HIV recency rate 0.005 and 95\% C.I. (0, 0.018). Since our classifier replaces the binary individual label of BCT with a probabilistic estimate, our estimated HIV recency rates and C.I.s are more realistic.

\begin{table}[!h]
\centering
\renewcommand{\thetable}{4.1}
\begin{tabular}{|c|c|c|c|c|c|}
\hline
\multicolumn{2}{|c|}{} & \multicolumn{2}{c|}{Our Classifier} & \multicolumn{2}{c|}{Raw BCT} \\ \hline
Level & n & Est & 95\% C.I. & Est & 95\% C.I. \\ \hline
F:15-24 & 89 & 0.155 & (0.110,0.204) & 0.070 & (0.016,0.139) \\ \hline
F:25-34 & 169 & 0.121 & (0.089,0.156) & 0.051 & (0.013,0.100) \\ \hline
F:35-44 & 114 & 0.083 & (0.054,0.124) & 0.005 & (0,0.018) \\ \hline
F:45-54 & 45 & 0.152 & (0.073,0.237) & 0.041 & (0,0.128) \\ \hline
F:55-64 & 21 & 0.275 & (0.089,0.445) & 0.090 & (0,0.286) \\ \hline
M:15-24 & 27 & 0.189 & (0.098,0.282) & 0.016 & (0,0.056) \\ \hline
M:25-34 & 95 & 0.126 & (0.091,0.168) & 0.030 & (0.005,0.072) \\ \hline
M:35-44 & 84 & 0.109 & (0.069,0.155) & 0.024 & (0.004,0.055) \\ \hline
M:45-54 & 42 & 0.085 & (0.038,0.166) & 0 & (0,0) \\ \hline
M:55-64 & 19 & 0.099 & (0.033,0.189) & 0.052 & (0,0.176) \\ \hline
\end{tabular}
\caption{Comparison of HIV recency rates between BCT and our classifier, by age-sex.  For each group (``Level"), we present the sample size (``n"), the estimate (``Est") and Bootstrap 95\% CI of our classifier, and the estimate (``Est") and Bootstrap 95\% CI of the raw BCT.}
\label{tab:compAgeSex}
\end{table}

We also present the summary of demographic and geographic, and present the example of gender-zone in Table \ref{tab:compGenZone}.  In addition to having zero HIV recency estimates (male Central-East and male Central-West) and zero lower bounds of 95\% CI (all levels except male Blantyre), the point estimates of BCT also vary dramatically. For instance, the estimated HIV recency rate among females in Central-West is 0.145, which is very high.  In contrast, the estimated HIV recency rate among males in Central-West is 0, and the estimated HIV recency rates among females in North and Central-East are 0.051 and 0.031 respectively. With our probabilistic classifier, we obtain more stable aggregated estimates which seems to be more reasonable than BCT estimates.


\begin{table}[!ht]
\centering
\renewcommand{\thetable}{4.2}
\begin{tabular}{|c|c|c|c|c|c|}
\hline
\multicolumn{2}{|c|}{} & \multicolumn{2}{c|}{Our Classifier} & \multicolumn{2}{c|}{Raw BCT} \\ \hline
Level & n & Est & 95\% C.I. & Est & 95\% C.I. \\ \hline
F:North & 33 & 0.139 & (0.074,0.218) & 0.051 & (0,0.143)  \\ \hline
F:CentralEast & 25 & 0.252 & (0.136,0.378) & 0.031 & (0,0.104) \\ \hline
F:CentralWest & 21 & 0.204 & (0.087,0.324) & 0.145 & (0,0.313)  \\ \hline
F:Lilongwe & 91 & 0.086 & (0.057,0.127) & 0.021 & (0,0.054)  \\ \hline
F:SouthEast & 55 & 0.157 & (0.097,0.222) & 0.045 & (0,0.114)  \\ \hline
F:SouthWest & 86 & 0.078 & (0.051,0.112) & 0.025 & (0,0.067) \\ \hline
F:Blantyre & 127 & 0.080 & (0.055,0.119) & 0.026 & (0,0.060)  \\ \hline
M:North & 17 & 0.101 & (0.041,0.177) & 0.040 & (0,0.145)  \\ \hline
M:CentralEast & 19 & 0.190 & (0.093,0.318) & 0 & (0,0) \\ \hline
M:CentralWest & 26 & 0.148 & (0.060,0.257) & 0 & (0,0)  \\ \hline
M:Lilongwe & 52 & 0.140 & (0.084,0.204) & 0.052 & (0,0.121)  \\ \hline
M:SouthEast & 35 & 0.095 & (0.054,0.147) & 0.025 & (0,0.090)  \\ \hline
M:SouthWest & 47 & 0.095 & (0.057,0.140) & 0.016 & (0,0.058)  \\ \hline
M:Blantyre & 71 & 0.113 & (0.075,0.160) & 0.050 & (0.009,0.104)  \\ \hline
\end{tabular}
\caption{Comparison of HIV recency rates between BCT and our classifier, by gender-zone.  For each group (``Level"), we present the sample size (``n"), the estimate (``Est") and Bootstrap 95\% CI of our classifier, and the estimate (``Est") and Bootstrap 95\% CI of the raw BCT.}
\label{tab:compGenZone}
\end{table}

\section{Simulation study}
\label{sec:Simulation}

In this section, we compare our semi-supervised learning method with a supervised learning method that only uses the contingency table data to train a model.  This alternative supervised method treats the contingency tables as repeated observations of categorical covariates, and fits a logistic regression on the table data.  Since the PHIA data does not have labels $Y_i$, we evaluate the predictions from the two methods by defining a true model and simulating contingency tables based on that model.

\subsection{Simulate the univariate ODn contingency table}

In this subsection, we simulate univariate ODn contingency tables. The same settings as the real data from Duong 2015 \citep{Duong2015} are used, including table covariate $T=ODn$, the same 6 table cells $\{C_1,\cdots,C_6\}$, the table sample size of 4733, and the table recency proportion with 994 $Y=1$ and 3739 $Y=0$.  We use the original 2015-2016 Malawi PHIA \citep{MPHIA2018} covariates and the estimated $\hat{P}(Y=1)=0.126$.

As the first step, we define a true model, assuming a logistic regression following Equation (\ref{eqn:logistic1}).  For the true model covariates, we select the table covariate, i.e., ODn, and possibly an auxiliary covariate, e.g., VL, and assign true coefficient values $\beta$ to these covariates.  Based on the external estimator $\hat{P}(Y=1)$, we compute the corresponding true intercept $\beta_0$.  With this true model, we also compute the true recency probabilities $P(Y_i=1|X_i;\beta, \beta_0)$ for Malawi PHIA individuals.

Next, we approximate the multinomial proportions of table data by using the empirical distribution in PHIA.  Based on the table recency proportion, which is different from the external estimator $\hat{P}(Y=1)$, we first compute the adjusted intercept $\beta_0^A$ and adjusted individual probabilities $P(Y_i=1|X_i;\beta, \beta_0^A)$.  We then compute $P(Y,T \in C_k; \beta)$ based on the empirical approximation in Equation (\ref{eqn:approx sim joint p_k}).  As we did in Equation \ref{eqn:approximation of p_k}, we group the PHIA individuals using the table covariate $T$ and table cells $C_k$, and use PHIA sampling weights $w_i$.
\begin{equation}
    P(Y=h,T \in C_k; \beta) \approx \sum_{i: T_{i} \in C_k}  P(Y_i=h|X_{i.};\beta, \beta_0^A)\times w_i
    \label{eqn:approx sim joint p_k}
\end{equation}

Finally, we simulate contingency tables from the multinomial distribution, using the approximated multinomial proportions $P(Y=h,T \in C_k; \beta)$ and table sample size.  We randomly generate 1000 contingency tables from this distribution.

\subsection{Alternative supervised learning method}

For each of the 1000 simulated contingency tables, we train a logistic regression model with categorical covariates.  To make the table data easier to fit, we first transform the simulated contingency table into standard format (observations vs features); the contingency table counts become repeated observations of binary response $Y$ and 6 ODn dummy variables (corresponding to the 6 table cell ranges).  Then, we fit a logistic regression on the transformed table dataset with binary response $Y$ and 5 dummy ODn variables.

For each simulated table, we apply our fitted categorical model on PHIA.  To match the format of the transformed table dataset, we transform the ODn variable in PHIA from continuous to the same 6 dummy variables (corresponding to the 6 table cell ranges).  We also adjust the intercept of the fitted categorical model based on external estimator $\hat{P}(Y=1)$.  Finally, we apply our fitted logistic model to predict the individual probabilities $\hat{P}_{alt}(Y_i=1)$ in PHIA.

\subsection{Results}

For each of the 1000 simulated contingency tables, we fit our method based on the procedure in Section \ref{subsection: combine data}, and predict the individual probabilities $\hat{P}_{our}(Y_i=1)$ in PHIA.  We manually select the model covariates to fit.

For each simulated table $r$, $r=1,\cdots,1000$, we evaluate the $n=705$ individual probability predictions using MAE (mean absolute error) from Equation (\ref{eqn:MAE}).  We compute MAE for our method and the alternative method, and obtain $MAE_r^{our}$ and $MAE_r^{alt}$.  For method comparison, we summarize the 1000 MAE values ($MAE_r^{our}$ or $MAE_r^{alt}$) by calculating their mean and SD.
\begin{equation}
    MAE_r^{pred} = \frac{1}{n} \sum_{i=1}^n |\hat{P}_{pred}(Y_i=1) - P(Y_i=1|X_i;\beta, \beta_0)|
    \label{eqn:MAE}
\end{equation}

For the first simulation, we assume a true logistic regression model in PHIA with covariate ODn and coefficient $\beta_{ODn}=-1$.  For our method, we manually fit ODn-only, and obtain .0032 (.0025) for the mean (SD) of the 1000 MAE values $MAE_r^{our}$.  The alternative method obtains .0371 (.0012) for the mean (SD) of the 1000 MAE values $MAE_r^{alt}$.  Our method outperforms the alternative method because our method utilizes more granularity from the ODn covariate from PHIA.  In logistic regression, our model can take ODn as a continuous covariate, while the alternative method takes ODn as a categorical dummy covariate, since the table only has discretized ODn.

For the second simulation, we assume a true logistic regression model in PHIA with covariates (ODn,VL) and coefficients $(\beta_{ODn},\beta_{VL})=(-1,1)$. For our method, we obtain .0168 (.0110) for the mean (SD) of the 1000 MAE values $MAE_r^{our}$ when we use (ODn,VL) as the covariate.  The alternative method obtains .0798 (.0005) for the mean (SD) of the 1000 MAE values $MAE_r^{alt}$. Our method outperforms the alternative method because our method can incorporate additional covariates (i.e., VL) from PHIA. The alternative method can only fit ODn because that is the only table covariate.


In both simulations, our method has the advantage of incorporating information from both PHIA and table, while the alternative method is limited to only using table data in the supervised learning.

\section{Discussion}
\label{sec:Conclusion}


After HIV prevalence has been stabilized, HIV incidence becomes the most important indicator for evaluating HIV/AIDS epidemics \citep{Hallett2011, WHO2020}. Unfortunately, HIV incidence, which is based on HIV recency status (recent vs long-term), is harder to estimate than HIV prevalence, which is based on HIV infection status (positive vs negative).
The innovative Population-based HIV Impact Assessment (PHIA) surveys are larger than previous national population-based household surveys, such as Demographic and Health Surveys (DHS) and AIS, and include additional biomarker tests for HIV antibody and HIV viral load \citep{MPHIA2018}. These biomarker tests offer the unique opportunity to detect recent HIV infection among the nationally-representative samples. In this article, we propose a statistical model for estimating the probability of being recently infected. With Malawi PHIA example, we demonstrate that the proposed model is more accurate for the individual level estimation and more appropriate for estimating HIV recency rates at aggregated levels than the current practice \,---\, the binary classification tree (BCT). 

This paper contributes to the literature by developing a semi-supervised learning method that combines the individual-level unlabeled data (PHIA) and the aggregated labeled data (cohort study). First, to compensate for the unknown individual labels in the primary data (PHIA), we borrow information from literature where the binary label is known at summary level, and assume that the relationship between X and Y is the same for the two datasets.  Second, to resolve the mismatch in covariate granularity and dimension, we further assume the two datasets have the same $P(X|T \in C_k)$. Note that we did not assume the same marginal distribution of covariates $X$ between data sources but the same conditional distribution given $T \in C_k$, and this assumption is reasonable when the range of $C_k$ is small.  Third, to resolve the different proportions of $Y=1$ (the primary dataset is representative and the contingency table is not), we calibrate the intercept $\beta_0$, and this follows standard case-control adjustment procedure \citep{Breslow1996}. Finally, we show that our proposed estimator for non-intercept coefficients are consistent and asymptotically normal and quantify the order of difference between the case-control adjustment estimator of the intercept and its true value. 



We apply the proposed method to predict the unknown binary label of individual HIV recency status in Malawi PHIA.  Model validation is challenging because we do not know the true recency status of individuals in PHIA.  To obtain datasets with known individual recency status, additional cohort studies need to be conducted in Malawi. In this article, we compare our proposed method with an alternative classification method \,---\, the BCT.  Our method outperforms the existing BCT method in the following aspects. First, as a data usage innovation, our method utilizes more information from covariates. By design, the BCT uses only 1 ODn threshold (1.5) and 1 VL threshold (1000). In contrast, our model takes ODn and VL as continuous covariates in logistic regression. Second, our classifier has the capability to incorporate additional covariates (e.g., self-reporting HIV positive/negative and tested for HIV status within a year), whereas the current BCT is limited to ODn and VL.  Third, the BCT can produce zero or outlier HIV incidence estimates (the proportion of people newly infected with HIV within a year) when the sample size is small for a sub-national region or a certain age-sex group, e.g., when no PHIA participant in the area is flagged as ``Recent".  Recent HIV infection is a relatively rare event among people currently living with HIV, and the small sample size can lead to large variability in proportion of PHIA individuals flagged as ``Recent".  We replace the binary prediction of HIV recency from the BCT with a probabilistic prediction, which naturally addresses the problem.  Fourth, our classifier appears more likely to identify truly recent individuals.  In addition to fitting high recency probability to BCT ``Recent" individuals (and agreeing with the BCT that those individuals should be classified as recent), our predictor also gives high recency probability to some BCT ``Long-Term" individuals.  These BCT ``Long-Term" individuals have (ODn,VL) values just outside the boundary of the BCT ``Recent" box, or have low ODn with low VL (which could still be recent since VL drops after initial infection \citep{Lavreys2002, Kahn1998}).  These individuals are also likely to self-report HIV negative, and this is a strong indicator of recent infection because they are unaware of their true HIV positive status.

PHIA surveys are being implemented in at least fifteen countries that are most-affected by the HIV/AIDS epidemic.  As of January 2021, two surveys (Eswatini, Mozambique) are in the pre-implementation phase, two are in data collection (Malawi, Uganda), and sixteen have completed fieldwork (Cameroon, Cote D’Ivoire, Eswatini, Ethiopia, Haiti, Kenya, Lesotho (two surveys), Malawi (used in this paper), Namibia, Rwanda, Tanzania, Uganda, Zambia, Zimbabwe (two surveys)). The proposed method is applicable to these PHIA surveys once public data become available, which is the case for four surveys (Eswatini, Malawi (used in this paper), Tanzania, Zambia) \citep{PHIA2020}.  With the additional PHIA surveys from other countries, we can further investigate the performance of the proposed model, such as the selection of covariates, the comparison with BCT classifications, and the validation of estimates provided for sub-national regions and age-sex groups. 



Beyond our specific application of HIV incidence estimation, our method could also be useful in other settings.  For example, in the context of social, behavioral and economic sciences, confidential individual-level data is often shared in the derivative form of summarized contingency tables \citep{Slavkovic2010}.  These contingency tables are seen readily available as official statistics products from the US Census, Bureau of Labor Statistics, and the Internal Revenue Service \citep{Fienberg2010, Yang2012, Barak2007}, to name a few. When the variable of interest is missing in an individual-level survey but present in the above contingency tables, one could use our proposed method to predict that variable for survey participants.

\section*{Acknowledgments}
Le Bao, Ben Sheng, Changcheng Li are supported by NIH/NIAID 5-R01-AI136664. Runze Li and Changcheng Li are supported by National Science Foundation, DMS 1820702, DMS 1953196 and DMS 2015539. 

\bibliographystyle{imsart-nameyear}
\bibliography{mainRef}

\appendix
\end{document}